\documentclass[fleqn,twoside]{article}
\usepackage{espcrc2}
\usepackage{epsfig}
\usepackage{graphicx}
\usepackage[figuresright]{rotating}

\newcommand{\AmS}{{\protect\the\textfont2
  A\kern-.1667em\lower.5ex\hbox{M}\kern-.125emS}}
\newcommand{\case}[2]{{\ensuremath{\textstyle\frac{#1}{#2}}}}
\title{Remark on the Theoretical Uncertainty in $B^0$-$\bar{B}^0$ Mixing 
	\vspace{-1.5cm}\begin{small}
	\begin{flushright}FERMILAB-Conf-02/212-T\\TRINLAT-02/06
	\end{flushright}\end{small} }

\author{Sin\'ead M.~Ryan\address[MCSD]{School of Mathematics, Trinity College, 
			    Dublin 2, Ireland} and
	Andreas S.~Kronfeld\address{Theoretical Physics Department, 
				       Fermi National Accelerator Laboratory,
				       Batavia, Illinois 60510, USA} }
\begin{document}
\begin{abstract}
	We re-examine the theoretical uncertainty in the Standard Model 
	expression for $B^0$-$\bar{B}^0$ mixing.
	We focus on lattice calculations of the ratio~$\xi$, needed to
	relate the oscillation frequency of $B^0_s$-$\bar{B}^0_s$
	mixing to $V_{td}$.
	We replace the usual linear chiral extrapolation with one that
	includes the logarithm that appears in chiral perturbation
	theory.
	We find a significant shift in the ratio~$\xi$, from the
	conventional $1.15\pm0.05$ to $\xi=1.32\pm0.10$.
\vspace{1pc}
\end{abstract}

\maketitle

\section{INTRODUCTION}
The Standard Model (SM) expression for $\bar{B}^0$--$B^0$ mixing, 
\begin{equation}
	\Delta m_q = \left(\frac{G_F^2m_W^2S_0}{16\pi^2m_{B_q}}\right) 
		|V_{tq}^*V_{tb}|^2 \eta_B {\cal M}_q, 
	\label{eq:Delta}
\end{equation}
where $q\in\{d,s\}$,
can be used to determine the CKM element $|V_{tq}|$. 
This result can be compared to other determinations to test the SM. In 
Eq.(\ref{eq:Delta}) the quantities in parenthesis are precisely known and 
$\Delta m_d=0.503\pm 0.006 \mbox{ps}^{-1}$~\cite{ref:Deltamd}. A measurement 
of $\Delta m_s$, at the percent level, is expected at the 
Tevetron~\cite{ref:Tevatron-mixing}. $|V_{tq}|$ is therefore dominated by 
QCD uncertainties in the hadronic matrix element,
${\cal M}_q = \langle\bar{B}_q^0| [\bar{b}\gamma^\mu (1-\gamma^5)q]
                            [\bar{b}\gamma_\mu (1-\gamma^5)q]|B_q^0\rangle$,
which is known to $\approx 20\%$ by lattice QCD~\cite{ref:Yamada02}. 
${\cal M}_q$ can be computed directly, or parameterised as 
${\cal M}_q = \frac{8}{3}m^2_{B_q}f^2_{B_q} B_{B_q}$ and reconstructed from 
separate determinations of $f_{B_q}$ and $B_{B_q}$. The latter approach is 
useful when studying the light quark mass dependence. A recent discussion of 
the methodologies and results is in Ref.~\cite{ref:Yamada02}. 

The percent-level accuracies of $\Delta m_s$ and $\Delta m_d$ make it
instructive to form the ratio
\begin{equation}
	\frac{\Delta m_s}{\Delta m_d} = 
		\left|\frac{V_{ts}}{V_{td}}\right|^2
		\frac{m_{B_s}}{m_{B_d}} \xi^2 
	\;\mbox{ with }\; \xi^2 = \frac{f_{B_s}^2 B_{B_s}}{f_{B_d}^2 B_{B_d}}
	\label{eq:ratio} .
\end{equation}
CKM unitarity implies $|V_{ts}|\approx |V_{cb}|$ and therefore 
$\delta |V_{td}| = \sqrt{ (\delta|V_{cb}|)^2 + (\delta\xi)^2 }$. 
Current uncertainty on $|V_{cb}|$ is 2-4\%, so the error in $|V_{td}|$ is 
largely due to the uncertainty in $\xi^2$.

Conventional wisdom holds that in the ratio $\xi^2$ many systemtic 
uncertainties cancel, implying a precise determination of $|V_{td}|$ is 
possible. 
However, typical lattice calculations use $0.5m_s\leq m_q\leq m_s$. To reach 
physical light quark masses the results are chirally extrapolated. The usual 
linear and quadratic fits, which may be reasonable in the region of lattice 
data, fail to account for the presence of chiral logarithms predicted by 
chiral perturbation theory~\cite{ref:SharpeZhang}. 
The effect of these logs was first explored by
JLQCD~\cite{ref:JLQCD-lat01} and included in the error budget of $\xi$ in 
Ref.~\cite{ref:Ryan01}. We argue that these logarithms change the 
extrapolation and therefore the value of $\xi$ in the chiral limit. 
We find a 
shift in $\xi$ from $1.15\pm 0.05$ to $1.32\pm 0.1$. 
Further details of this work are in Ref.~\cite{ref:KronfeldRyan}. 
        \vspace{-2ex}
        \begin{figure}[h]
        \begin{center}
        \epsfxsize=7.5cm\epsfbox{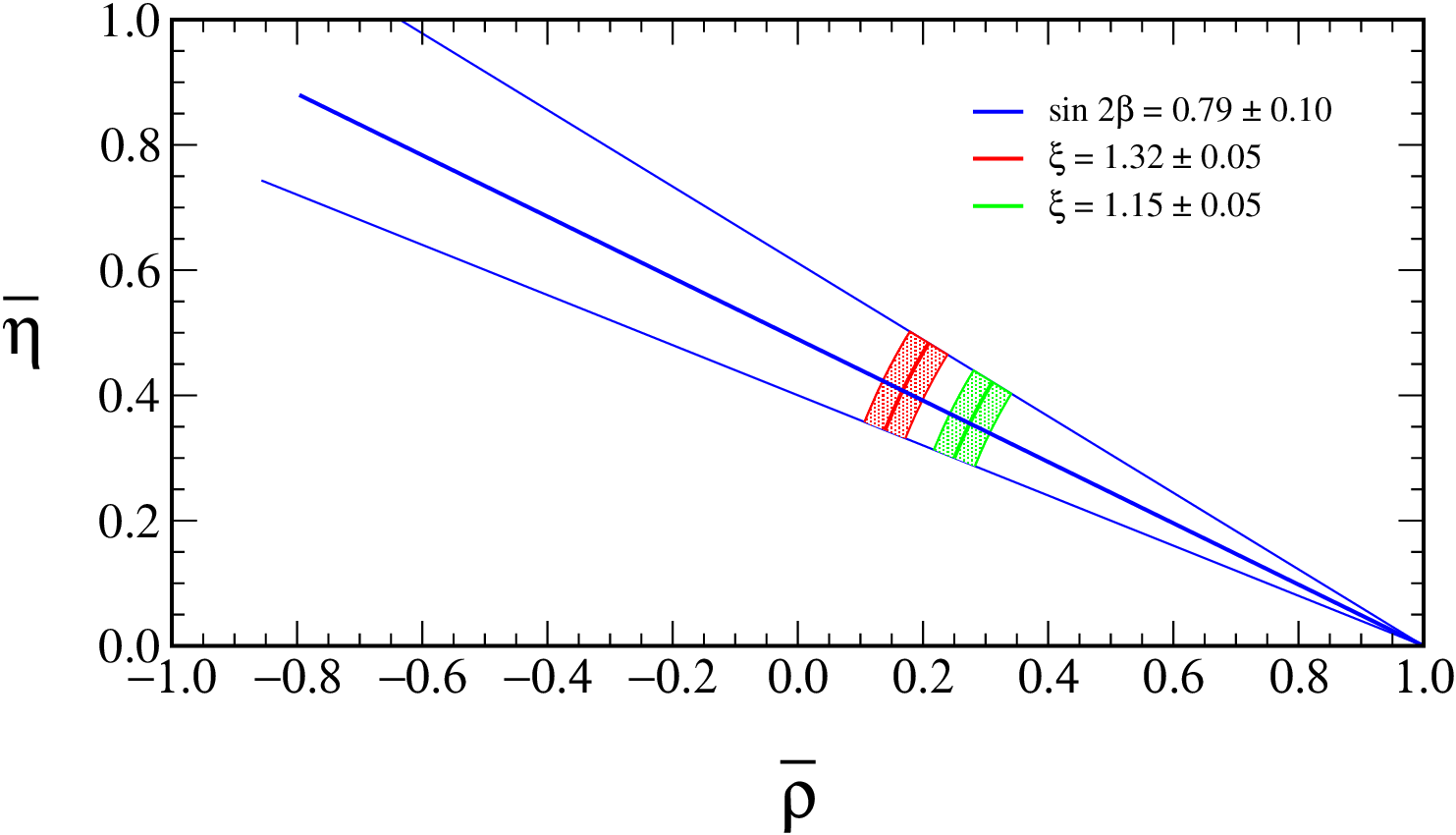}
        \vspace{-1cm}
        \caption{A sketch of the constraints on the apex of the
	unitarity triangle with $\sin 2\beta=0.79\pm0.10$,
	$\Delta m_s=20~\textrm{ps}^{-1}$ and
	$\xi=1.32\pm0.05$ or $1.15\pm0.05$.}
        \label{fig:RhoEta}\vspace{-0.05truein}
        \end{center}
        \vspace{-4ex}
        \end{figure}
Fig.~\ref{fig:RhoEta} shows how 
$\sin 2\beta$ and $\Delta m_s/\Delta m_d$ combine to constrain the apex of the 
unitarity triangle. The effect of a shift in $\xi$ is highlighted. 
\section{CHIRAL LOGS AND $\mathbf{\xi}$}
The scales between $1/m_s$ and $1/m_d$ are best described by chiral 
perturbation theory. Neglecting $1/m_b$ corrections
\begin{eqnarray}
	\sqrt{m_{B_q}} f_{B_q} & = & \Phi \left[1 + \Delta f_q \right],\\
	               B_{B_q} & = &   B  \left[1 + \Delta B_q \right],
\end{eqnarray}
where $\Delta f_q$ and $\Delta B_q$ contain the logarithms and are given in 
Ref.~\cite{ref:SharpeZhang}. 
The expressions for $\Delta f_q$ and $\Delta B_q$ yield
\begin{eqnarray}
 \xi_f-1\hspace{-0.3cm}&=&\hspace{-0.3cm}\Delta f_s - \Delta f_d \nonumber\\
              &=&\hspace{-0.3cm} (m_K^2 - m_\pi^2) f_2(\mu) -
        \frac{1+3g^2}{(4\pi f)^2} \bigg( 
        \case{1}{2} m_K^2    \ln\case{m_K^2}{\mu^2} \nonumber \\ 
	      &&+ 
        \case{1}{4} m_\eta^2 \ln\case{m_\eta^2}{\mu^2} -
        \case{3}{4}m_\pi^2 \ln\case{m_\pi^2}{\mu^2} \bigg),
					\hspace{-0.2cm}\label{eq:xifchi}\\
 \xi_B^2- 1\hspace{-0.3cm}&=&\hspace{-0.3cm}\Delta B_s-\Delta B_d \nonumber\\
           &=&\hspace{-0.3cm} (m_K^2 - m_\pi^2) B_2(\mu) -
        \frac{1-3g^2}{(4\pi f)^2} \bigg(
        \case{1}{2} m_\eta^2 \ln\case{m_\eta^2}{\mu^2} \nonumber \\  
			   &&-
        \case{1}{2} m_\pi^2  \ln\case{m_\pi^2}{\mu^2} \bigg) .
\end{eqnarray}
$f_2(\mu )$ and $B_2(\mu )$ are low-energy constants describing dynamics at scales $< \mu^{-1}$, $f$ is the pion decay constant and $g^2$ is the 
$B^\ast B\pi$ coupling. This has recently been determined by CLEO from a 
measurement of the $D^\ast$ width. They find 
$g^2_{D^\ast D\pi} = 0.35$~\cite{ref:CLEOgDD*pi} and by heavy quark symmetry, 
$g^2_{B^\ast B\pi} = 0.35$, to a good approximation. Thus, while 
the effect of the chiral log in $B_d$ is small, since $(1-3g^2)=-0.05$, it 
may be significant for $f_{B_d}$, where $(1+3g^2)=2.05$.

\begin{figure}[h]
        \begin{center}
        \epsfxsize=7.5cm\epsfbox{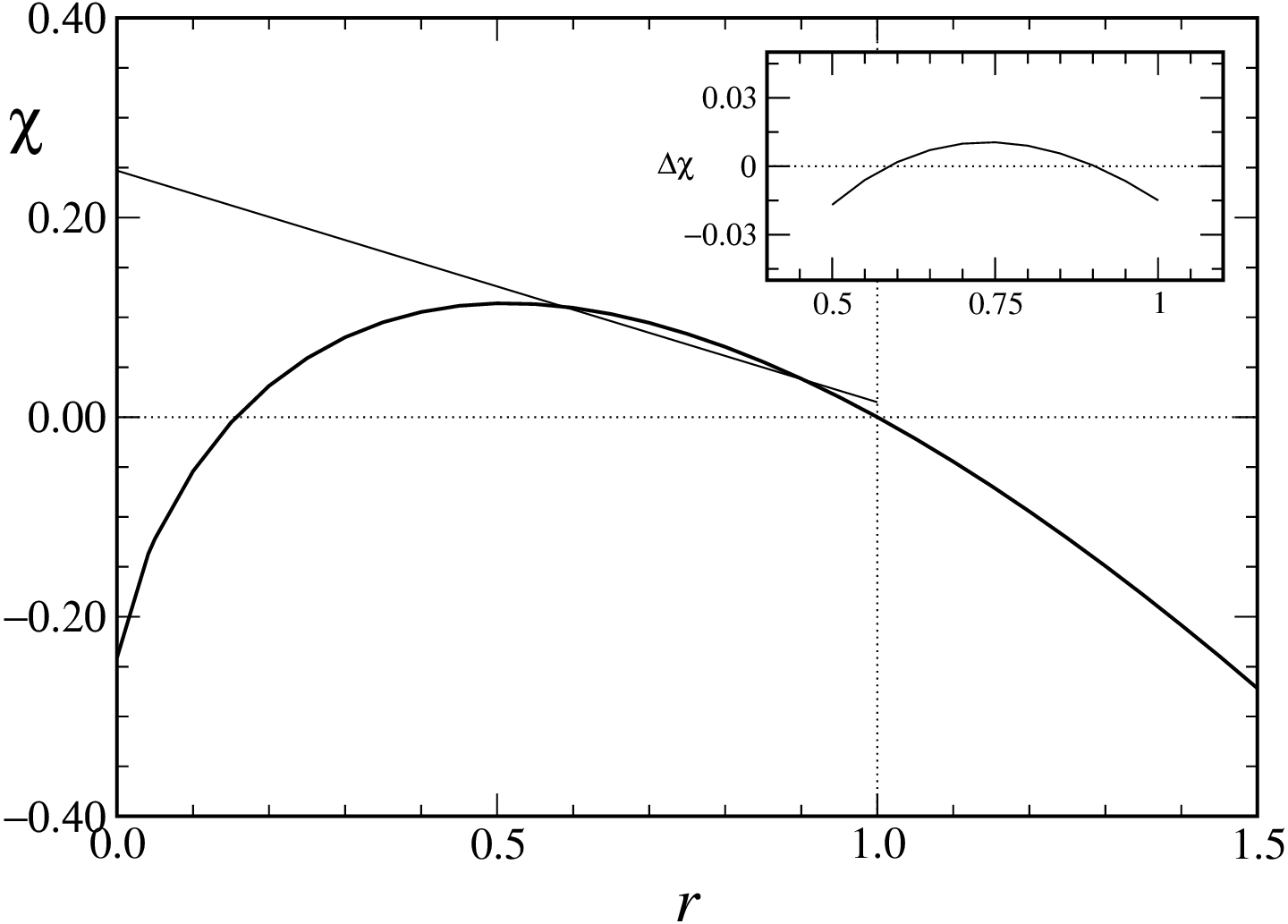}
        \vspace{-1cm}
        \caption{The chiral log $\chi(r)$
	varying the mass ratio $r=m_{qq}^2/m_{ss}^2=m_q/m_s$,
	compared with a straight line fit for $0.5\le r\le 1.0$.
	The difference between the curve and the fit is shown in
	the inset.}
        \label{fig:chi}\vspace{-0.05truein}
        \end{center}
        \vspace{-3ex}
        \end{figure}

Therefore, we refine our discussion and consider $\xi_f = f_{B_s}/f_{B_d}$.
Further details of the chiral log in $B_d$ and the effect of varying $g^2$
are in Ref.~\cite{ref:KronfeldRyan}. 
To study the light quark mass dependence of $\xi_f$, Eq. (\ref{eq:xifchi}) is
rewritten using the Gell-Mann--Okubo relations, as
\begin{eqnarray}
  \xi_f(r) - 1\hspace{-0.2cm}&=&\hspace{-0.2cm}m_{ss}^2(1 - r) 
	\bigg\{ \case{1}{2}f_2(\mu)\nonumber\\
	&-&\left. \frac{1+3g^2}{(4\pi f)^2} \left[\frac{5}{12}
		       \ln(\case{m_{ss}^2}{\mu^2}) + l(r) \right]\right\},
	\label{eq:xifr}
\end{eqnarray}
where $m_{qq}^2=rm_{ss}^2$ and 
\begin{eqnarray}
l(r)=\frac{1}{1-r}&&\hspace{-.6cm}\left[ 
       \frac{1+r}{4}  \ln\left(\frac{1+r}{2}\right)\right.\nonumber\\ 
   &&\hspace{-.8cm}+\left.\frac{2+r}{12} \ln\left(\frac{2+r}{3}\right) - 
	\frac{3r}{4}   \ln(r) \right].
	\label{eq:chilog}
\end{eqnarray}
Fig.~\ref{fig:chi} shows the function $\chi(r)=(1-r)l(r)$ which contains 
the chiral logarithms.
        \vspace{-2ex}
Typical lattice calculations are performed at $0.5\leq r\leq1.0$ and 
extrapolated to $r_d\approx 1/25$, corresponding to the down quark. 
Fig.~\ref{fig:chi} shows that the difference between a linear fit and one
including chiral logarithms is easily masked by statistical error in the 
region where data are available. However, the effect of the logarithms is 
significant at smaller $r$ values and is not reproduced by a linear fit. 

\begin{figure}[h]
        \begin{center}
        \epsfxsize=7.5cm\epsfbox{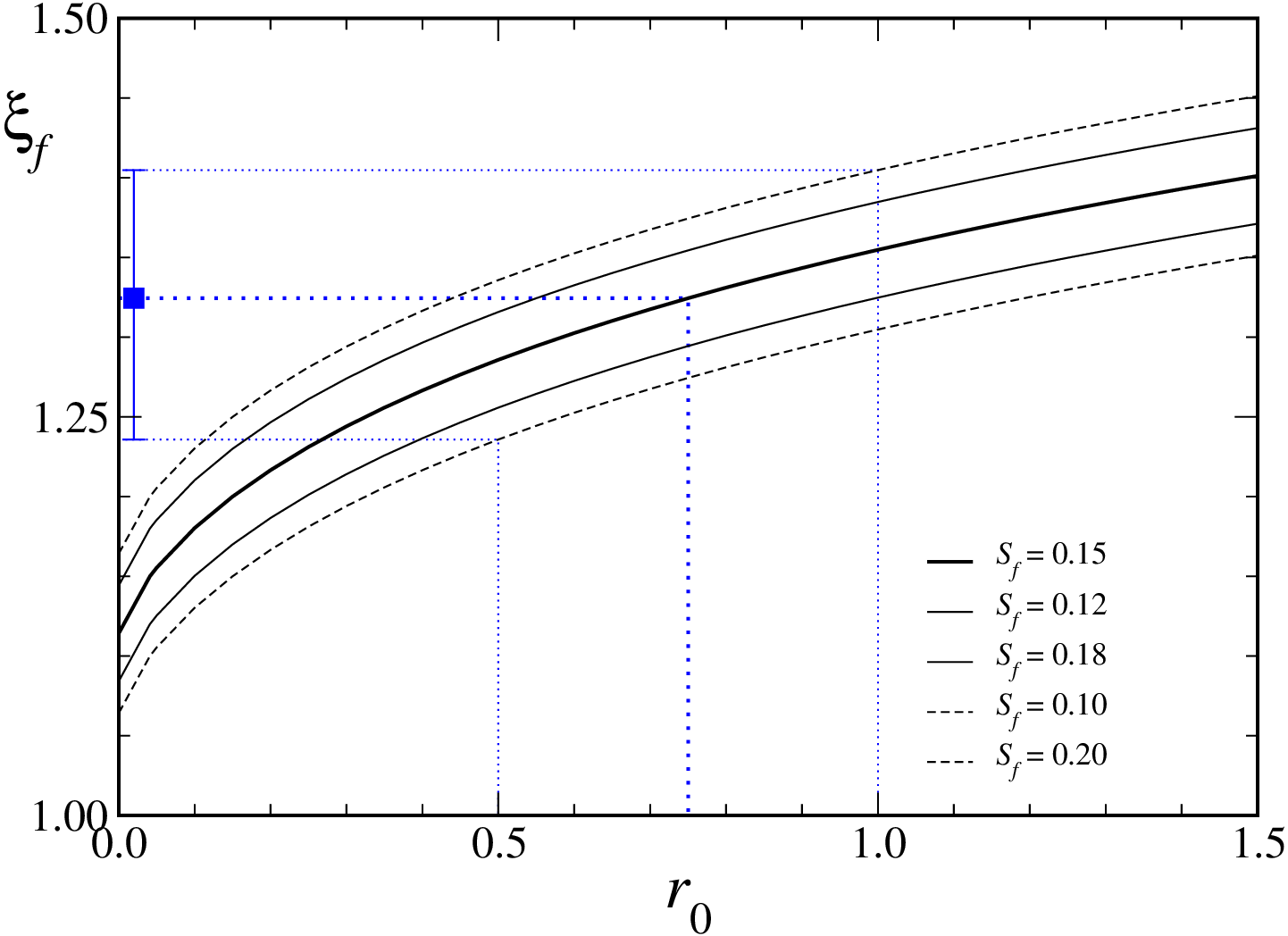}
        \vspace{-1cm}
        \caption{The dependence of $\xi_f$ on $r_0$. $(1-r_d)S_f=0.15\pm 0.05, 
		 m_{ss}^2=2(m_K^2-m_\pi^2), r=r_d=m_\pi^2/m_{ss}^2$.} 
        \label{fig:xi-r0}\vspace{-0.05truein}
        \end{center}
        \vspace{-3.5ex}
        \end{figure}

To include these logarithms in the chiral extrapolation of $\xi_f$ (or 
$\xi_f-1$) the low-energy constant $f_2(\mu )$ is required. 
This can be extracted from current lattice data
\footnote{Note that the dependence on the scale $\mu$ cancels in the total.}. 
The usual, linear functional form applied to $\xi_f-1$ is 
\begin{equation}
\xi_f(r) = (1-r)S_f \label{eqn:xi-linear}.
\end{equation}
Assuming this is sensible for $r=r_0\sim 1$, as indicated by 
Fig.~\ref{fig:chi}, and by equating Eqs. (\ref{eq:xifr}) and 
(\ref{eqn:xi-linear})
\begin{eqnarray}
	\frac{m_{ss}^2}{2}f_2(\mu)\hspace{-.8cm}&&= S_f \nonumber \\
				   &+& m_{ss}^2
		\frac{1+3g^2}{(4\pi f)^2} \left[\frac{5}{12}
			\ln(\case{m_{ss}^2}{\mu^2}) + l(r_0) \right],
	\label{eq:f2S}
\end{eqnarray}
which plugged into Eq. (\ref{eq:xifr}) gives
\begin{eqnarray}
	\xi_f(r) - 1 = &&\hspace{-.8cm}(1 - r) \bigg\{S_f \nonumber\\
	 	       &+&\hspace{-.1cm}\left. m_{ss}^2 
	\frac{1+3g^2}{(4\pi f)^2} \left[ l(r_0) - l(r) \right] \right\}.
	\label{eq:xiffinal}
\end{eqnarray}
Knowing $g^2$ from experiment and $S_f$ from 
lattice calculations gives a new handle on $\xi_f$. To evaluate 
$\xi_f(r)-1$, $f=130$ MeV, $g^2=0.35$ and 
$S_f=0.15\pm 0.05$~\cite{ref:Ryan01} are used. 
The parameter $r_0$ is a remaining uncertainty. Fig.~\ref{fig:xi-r0} shows 
the dependence of $\xi_f$ on $r_0$. 
        \vspace{-2.0ex}
Fig.~\ref{fig:xi-r0} yields a value of $\xi_f=1.32\pm0.08$. The conservative 
error attached to $S_f$ leads to the larger then usual error on $\xi_f$. A 
similar analysis of $\xi_B$ is described in Ref.~\cite{ref:KronfeldRyan}, 
giving $\xi_B=0.998\pm0.025$. The combination leads to
\begin{equation}
\xi = 1.32\pm 0.10 .\label{eqn:xi}
\end{equation}
%
\section{CONCLUSIONS}
The importance of reliable chiral extrapolations has become more and more 
widely appreciated~\cite{ref:panel-discussion}.
This is especially true in the quenched approximation where many other 
systematic errors have been controlled, leaving chiral extrapolation as 
the major uncertainty~\cite{ref:FNALB2piref:JunkoB2pi}.
To date, most lattice calculations of $\xi$ have relied on linear 
(or quadratic) fits. A combination of the new CLEO value of the coupling with 
lattice data 
for $f_2(\mu )$ allows a determination of $\xi$ incorporating chiral 
logarithm effects. Reducing the uncertainty in $\xi$ is possible by 
designing lattice calculations to determine the low-energy constants. 

Ultimately, unquenched lattice calculations at light quark masses should 
``see'' the chiral logarithms. 
Until then, the central value and error assigned to $\xi$ 
should reflect the uncertainty. 

\end{document}